\documentclass[10pt]{iopart}
\usepackage{graphicx}
\usepackage{amsfonts}
\usepackage{amsmath}
\usepackage{amssymb}
\usepackage{hyperref}
\usepackage{cite}
\usepackage{bbding}
\usepackage{makecell}
\usepackage{mathtools, cuted}

\usepackage{iopams}  
\begin{document}

\title[Quantum dot on a plasma-facing surface]{Quantum dot on a plasma-facing microparticle surface: Thermal balance}

\author{M.Y. Pustylnik}
\address{Institut f\"ur Materialphysik im Weltraum, Deutsches Zentrum f\"{u}r Luft- und Raumfahrt (DLR), Linder H\"ohe, 51147 Cologne, Germany}
\ead{mikhail.pustylnik@dlr.de}
\vspace{10pt}
\begin{indented}
\item[]
\end{indented}

\begin{abstract}
Semiconductor nanocrystals, quantum dots, are known to exhibit the quantum-confined Stark effect which reveals itself in the shift of their photoluminescence spectra in response to external electric field.
It was, therefore, proposed to use quantum dots deposited on the microparticle surface for the optical measurement of the charge acquired by the microparticles in low-temperature plasmas.  
Thermal balance of a quantum dot residing on the surface of a microparticle immersed in a plasma is considered in this work.
It is shown for typical plasma parameters that under periodically pulsed plasma conditions, the spectral shift of the photoluminescence of the quantum dot caused by the oscillations of its temperature becomes undetectable at the effective thermal flux characterizing the thermal contact between the quantum dot and the microparticle $\sim 10^9$~s$^{-1}$.
Under these conditions, the entire spectral shift observed during the period of plasma pulsing should be attributed to the quantum-confined Stark effect due to the microparticle charge.
Lower-boundary estimate for the effective thermal flux for the direct contact between the quantum dot and the microparticle is $\sim 10^{12}$~s$^{-1}$.
\end{abstract}
\maketitle
\ioptwocol
\hypersetup{breaklinks=true}

%
%
%
%
%
\section{Introduction}
\label{Sec:Intro}
{\it In-situ} measurement of the charge accumulated on small particles is a problem in many areas of science and technology, e.g., in the investigations of dust-growing reactive plasmas \cite{Watanabe1993, Fridman1996, Denysenko2020}, in the usage of complex (dusty) plasmas as atomistic model systems for the investigations of classical condensed matter phenomena \cite{RevModPhys.81.1353, ILMRbook}, in the investigations of Lunar environment \cite{Popel2018}, in the usage of microparticles as small optically manipulated plasma probes \cite{Schneider2018} as well as in the investigations of triboelectric charging \cite{KWETKUS1998, Sharma2008, Carter2017}.
Traditionally, the charge of dust particles is measured using dynamical methods \cite{Trottenberg1995, Fortov2001, Fortov2004, Khrapak2005, Nosenko2018, Antonova2019} which have known disadvantages such as necessity for (often not easily verifiable) assumptions on the forces acting on dust particles and limited spatiotemporal resolution.
\\ \indent
Optical methods of dust particle charge measurement are therefore of great interest.
Significant theoretical efforts have been undertaken to explore the possibilities of optical detection of the dust charge.
The surplus electrons modify the dielectric permittivity of the dust particle material or surface conductivity of the microparticle and therefore affect some of the spectral features of light scattering.
In \cite{Heinisch2012, Heinisch2013}, excitonic resonance was considered, whereas \cite{Vladimirov2016, Vladimirov2017} investigated the possibility of usage of the surface plasmon resonance.
Very recently, the possibility of the measurement of charges of silica nanoparticles in plasmas using Fourier-transform infrared spectroscopy of the excitonic resonance was experimentally demonstrated \cite{PhysRevE.104.045208}.
\\ \indent
Also, very recently, it was shown that semiconductor nanocrystals, quantum dots (QDs) \cite{Ekimov1981, Reed1988, Gaponenko1998}, deposited on a large flat surface are sensitive to the charge on this surface \cite{Marvi2021} due to the quantum-confined Stark effect \cite{Ekimov1990}:
Spectrum of their photoluminescence experiences red shift in reaction to the local electric field.
In \cite{Pustylnik2021}, it was proposed to design a charge microsensor by attaching the QDs to the surface of a micrometer-sized spherical particle.
The calculations showed that under typical conditions of dusty plasma experiments, such a microsensor can exhibit Stark shifts of the order of fractions of a nanometer.
The advantage of using the QDs compared to excitonic resonance is that the measurement can be performed in the visible spectral range.
\\ \indent
Experiments of \cite{Marvi2021} have shown that exposure of QDs to the fluxes of charged particles is connected not only with the surplus-charge-induced Stark shift, but also with the thermal shift of the photoluminescence spectrum \cite{Walker2003, Valerini2005, Albahrani2018}.
Charged particles bring their kinetic as well as potential energy to the plasma-facing surface, which is cooled by neutral gas as well as by thermal radiation \cite{Swinkels2000, Khrapak2006, Maurer2008, Polyakov2021}.
Heating is, therefore, along with charging, an unavoidable consequence of the exposure of the surface to almost any ionized medium.
In the scope of usage of QDs for the surface charge measurement, the spectral shifts caused by these two phenomena have to be distinguished.
\\ \indent
Charging of a plasma-facing surface usually occurs much faster than heating. 
On this basis, the so-called ``fast'' red shift in \cite{Marvi2021} was attributed to the Stark effect.
In accord with that, in \cite{Pustylnik2021}, periodic pulsing of the plasma (on the timescales between those of charging and heating) was suggested to distinguish between the thermal and electrostatic effects on the QD photoluminescence spectrum.
The weak point of this approach is that this is, in fact, not the surface temperature, but rather the temperature of the QDs which determines the thermal spectral shift.
QDs on the plasma-facing surface would not necessarily acquire the same equilibrium temperature as the substrate surface itself.
Also, their thermal inertia is much less than that of the substrate they are sitting on.
These two issues raise the question of the thermal contact between the QD and the surface it is attached to:
Is it sufficient to suppose the QD being thermally bound to the surface?
\\ \indent
This is the question we are addressing in the present paper.
We consider the thermal balance for the spherical micrometer-sized particle immersed in a plasma under typical experimental conditions.
Along with that, we consider the thermal balance for the QD placed on the surface of this particle.
Steady-state as well as periodically pulsed plasma conditions are considered.
Based on the solutions of thermal balance equations, we formulate the requirements for the thermal contact between the QD and the particle surface which should be used as one of the input parameters for the design of the charge microsensor.
\\ \indent
Our paper is organized as follows:
In \Sref{Sec:Mod}, we describe the thermal model for the QD and the microparticle, postulate the assumptions and formulate equations both for steady-state and pulsed plasma cases.
In \Sref{Sec:Res}, we show the results of the calculations.
\Sref{Sec:Disc} contains the discussion of the results, in particular, of the following issues:
Character of the QD temperature oscillations, their experimental detectability, cooling regimes of the QD and, finally, quantitative requirements on the thermal contact between the QD and the microparticle.
In \Sref{Sec:Conc}, we draw the conclusions.
\section{Model}
\label{Sec:Mod}
\begin{figure}[t!]
\includegraphics[width=8cm]{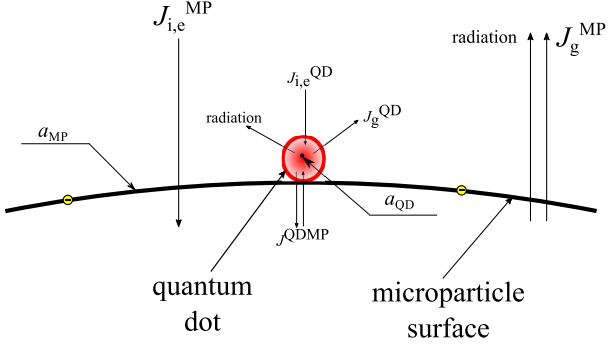}
\caption{
Schematic representation of the thermal balance for a plasma-facing microparticle surface and quantum dot residing on it.
The surfaces are heated by the fluxes of ions and elecrons $J_{\rm i,e}^{\rm MP, QD}$ and cooled by the fluxes of neutral atoms $J_{\rm g}^{\rm MP, QD}$ and thermal radiation.
Flux $J^{\rm QDMP}$ characterizes the thermal contact between the microparticle surface and quantum dot.}
\label{Fig:Sch}
\end{figure}
In order to calculate the temperatures of a microparticle and of a QD residing on its surface, we consider the thermal balance for both bodies taking into account the heat fluxes from the plasma to their surfaces and visa versa as well as heat exchange between them.
The fluxes taken into account in the model are schematically shown in \Fref{Fig:Sch}.
Along with \cite{Swinkels2000, Akdim2003, Khrapak2006}, we suppose that energy is brought to plasma-facing  surfaces by charged species and is taken away by neutral gas and thermal radiation.
\\ \indent
We first formulate and discuss the model for the steady-state plasma.
Afterwards, we discuss the assumptions and write the equations for the case of periodically pulsed plasma.
\subsection{Steady-state plasma}
\subsubsection{Electron, ion and neutral fluxes}
We start the formulation of our model with the charged-particle fluxes onto the surface of the microparticle.
We suppose a microparticle to be immersed into an isotropic plasma with electron temperature $T_{\rm e}$, density $n$, and ions having the temperature of neutral gas $T_{\rm g}$.
Here, we follow the approach of \cite{Lampe2003} in which the orbit-motion-limited $J_{\rm e}^{\rm MP}$ current is implied for electrons, whereas for ions, the collisions with neutrals in the vicinity of the microparticle have to be considered.
The electron current is then expressed as follows:
\begin{equation}
J_{\rm e}^{\rm MP}=\pi a_{\rm MP}^2nv_{\rm eth}\exp{\left(-\frac{e\phi}{k_{\rm B}T_{\rm e}}\right)},
\label{Eq:JeMP}
\end{equation}
where $a_{\rm MP}$ is the microparticle radius, $v_{\rm eth}=\sqrt{8k_{\rm B}T_{\rm e}/\pi m_{\rm e}}$ is the electron thermal velocity and $\phi$ is the absolute value of microparticle surface potential (which is normally negative).
The expression for the ion current reads:
\begin{equation}
J_{\rm i}^{\rm MP}=\pi a_{\rm MP}^2nv_{\rm ith}\left(1+\frac{e\phi}{k_{\rm B}T_{\rm g}}+\frac{R^3}{a_{\rm MP}^2l_{\rm i}}\right),
\label{Eq:JiMP}
\end{equation}
where $R$ denotes the radius of a sphere around a microparticle, inside which the absolute value of the electrostatic potential is below $T_{\rm g}$,  $v_{\rm ith}=\sqrt{8k_{\rm B}T_{\rm g}/\pi M}$ ($M$ is the ion mass) is the ion thermal velocity and $l_{\rm i}$ is the ion mean free path.
In case of Yukawa potential around the microparticle,  $R$ should satisfy the equation:
\begin{equation}
\frac{e\phi}{k_{\rm B}T_{\rm g}}\exp{\left(-\frac{R}{\lambda_{\rm D}}\right)}=\frac{R}{a_{\rm MP}},
\end{equation}
where $\lambda_{\rm D}$ is the Debye length.
\\ \indent
For the neutral flux onto the microparticle, we use the expression from \cite{Khrapak2006}:
\begin{equation}
J_{\rm g}^{\rm MP}=2\pi a_{\rm MP}^2v_{\rm gth}\frac{p}{k_{\rm B}T_{\rm g}},
\label{Eq:JgMP}
\end{equation}
where $p$ is the neutral gas pressure and $v_{\rm gth} \equiv v_{\rm ith}$ is the thermal velocity of neutral atoms.
\\ \indent
For the microparticle, the fluxes of the plasma species are isotropically collected over its entire surface area.
This will not be true for the QD.
Therefore, the respective currents will read:
\begin{equation}
	J_{\rm e,i,g}^{\rm QD} = \alpha_{\rm e,i,g}J_{\rm e,i,g}^{\rm MP}\left(\frac{a_{\rm QD}}{a_{\rm MP}}\right)^2,
	\label{Eq:JeiQD}
\end{equation}
where $a_{\rm QD}$ is the QD radius and $\alpha_{\rm e,i,g}$ is the fraction of the total QD surface area over which the electron, ion and neutral fluxes are respectively corrected.
For the cahrged species, Equation~\ref{Eq:JeiQD} implies that the surface of the QD is under the same electrostatic potential as that of the microparticle.
\subsubsection{Thermal radiation}
Characteristic wavelength of thermal radiation $\lambda=hc/k_{\rm B}T_{\rm g}$ is about $50$~$\mu$m at room temperature.
Since $a_{\rm MP,QD}<\lambda$, the intensity of the thermal radiation of both the microparticle and the QD will be lower than that of the black body.
We assume the thermal radiation intensity of the microparticle and the QD $I_{\rm rad}^{\rm MP,QD}=\epsilon_{\rm MP, QD}\sigma T_{\rm MP,QD}^4$, where $\epsilon_{\rm MP, QD}$ denote the emissivities of the microparticle and the QD, respectively.
According to \cite{Rosenberg2007}, the emissivity depends on the size of a small object as well as on the complex refractive index of its material.
\subsubsection{Charging}
We suppose that the QD and the microparticle surface can freely exchange charge.
In addition to that, it was shown in \cite{Pustylnik2021} that the typical average distances between the electrons on the microparticle surface is much larger than the QD size.
The QD, although it collects the charged particles, will, consequently, most of the time stay uncharged.
We will therefore not consider its charging process separately and will only include charging of the entire microparticle described by the equation
\begin{equation}
\frac{\partial \phi}{\partial t}=\frac{e}{4\pi\epsilon_0a_{\rm MP}}\left(J_{\rm e}^{\rm MP}-J_{\rm i}^{\rm MP}\right).
\label{Eq:Ch}
\end{equation}
\subsubsection{Thermal balance}
When considering the thermal balance for both the microparticle and the QD, we suppose that every ion impinging one of these surfaces deposits its entire kinetic energy $e\phi$ as well as its entire potential energy $E_{\rm i}$ released due to its recombination on the respective surface.
In the same way, every electron will deposit the energy $k_{\rm B}T_{\rm e}$.
\\ \indent
Supposing $\xi_{\rm MP, QD}$ to be the accommodation coefficients and $T_{\rm MP, QD}$ the temperatures of microparticle and QD surfaces, respectively, every neutral atom will carry away the energy $\xi_{\rm MP, QD}k_{\rm B}\left(T_{\rm MP, QD}-T_{\rm g}\right)$ from the respective surface.
\\ \indent
We assume the plasma to be transparent for the thermal radiation.
The reactor walls are supposed to stay at the temperature of the neutral gas.
Therefore, the microparticle and the QD will absorb the thermal radiation with the temperature $T_{\rm g}$ and, at the same time, emit radiation with the temperature $T_{\rm MP,QD}$.
\\ \indent
The thermal balance equations can then be written as\\
\begin{strip}
\begin{eqnarray}
C_{\rm MP}m_{\rm MP}\frac{\partial T_{\rm MP}}{\partial t}=J_{\rm i}^{\rm MP}\left(E_{\rm ion}+e\phi\right)+J_{\rm e}^{\rm MP}k_{\rm B}T_{\rm e}-\xi_{\rm MP}J_{\rm g}^{\rm MP}k_{\rm B}\left(T_{\rm MP}-T_{\rm g}\right)-4\pi a_{\rm MP}^2\epsilon_{\rm MP}\sigma\left(T_{\rm MP}^4-T_{\rm g}^4\right) 
\label{Eq:TBMP}\\
\begin{aligned}
C_{\rm QD}m_{\rm QD}\frac{\partial T_{\rm QD}}{\partial t}=J_{\rm i}^{\rm QD}\left(E_{\rm ion}+e\phi\right)+J_{\rm e}^{\rm QD}k_{\rm B}T_{\rm e}-\xi_{\rm QD}J_{\rm g}^{\rm QD}k_{\rm B}\left(T_{\rm QD}-T_{\rm g}\right)&-4\pi a_{\rm MP}^2\epsilon_{\rm QD}\sigma\left(T_{\rm QD}^4-T_{\rm g}^4\right)-\\
						&-J^{\rm QDMP}k_{\rm B}\left(T_{\rm QD}-T_{\rm MP}\right),
\end{aligned}
\label{Eq:TBQD}
\end{eqnarray}
\end{strip}
\noindent
where $C_{\rm MP, QD}$ are the heat capacities of the microparticle and QD material, respectively, $m_{\rm MP, QD}$ are the masses of the microparticle and QD, respectively, and $J^{\rm QDMP}$ is the effective flux characterizing the thermal contact between the QD and the microparticle.
Equilibrium values of $T_{\rm MP, QD}$ will be determined by zeroing the derivatives in Equations \eref{Eq:Ch}, \eref{Eq:TBMP} and \eref{Eq:TBQD}.
\subsection{Periodically pulsed plasma}
\label{Sec:PPP}
We are however mostly interested not in the steady-state temperatures, but in the oscillations of $T_{\rm QD}$ in response to pulsing the plasma.
As already mentioned above, pulsing of the plasma was suggested in \cite{Pustylnik2021}, as a possible measure to distinguish between the thermal and charge-induced Stark shift of the photoluminescence spectrum of the plasma-facing QDs.
Thermal contact of the QD and the microparticle (represented by the unknown effective flux $J^{\rm QDMP}$ in \Eref{Eq:TBQD}) should provide the QD sufficient thermal inertia making the optical charge measurements possible.
\\ \indent
If the plasma is periodically pulsed, it periodically goes through reignition and afterglow phases, which are very hard to properly model.
Also coefficients $\alpha_{\rm e,i,g}$, $\xi_{\rm MP,QD}$ and $\epsilon_{\rm MP,QD}$ entering Equations~\eref{Eq:TBMP} and \eref{Eq:TBQD} cannot be easily calculated from first principles.
Therefore, when considering our problem, we will use certain simplifications which should target to increase the thermal contrast (i) between the afterglow and reignition conditions as well as (ii) between the microparticle and the QD.
\subsubsection{Heat fluxes}
\label{Sec:HF}
Afterglow is quite a long process \cite{Celik2012}.
Even milliseconds after switching off the electrical power, significant densities of the charged species remain in the reactor volume.
Therefore, in a real afterglow, the microparticle and QD surfaces will continue collecting heat.
On the other hand, after the plasma reignition, the denstities of the charged species and, consequently, the heat fluxes to the microparticle and QD surfaces also do not immediately reach the steady-state values.
\\ \indent
Taking that into account, in our model, we will suppose that the heat fluxes immediately vanish once the plasma is switched off and immediately acquire steady-state values once the plasma is switched on.
This eliminates the necessity to solve charging \Eref{Eq:Ch} since we will always assume steady state microparticle surface potential.
Since cooling fluxes and thermal contact between the microparticle and the QD are independent of the plasma, they will be present in the model at all times. 
\subsubsection{Surface conditions of the microparticle and the QD}
\label{Sec:Surf}
Relevant for our problem surface conditions of the microparticle and the QD are represented by the emissivities $\epsilon_{\rm MP,QD}$ which are related to thermal radiation as well as by the accommodation coefficients $\xi_{\rm MP,QD}$ and coefficients $\alpha_{\rm e,i,g}$ which are related to the collection of plasma species.
\\ \indent
According to \cite{Rosenberg2007}, $\epsilon_{\rm MP,QD}~\sim x_{\rm MP,QD}=2\pi a_{\rm MP,QD}/\lambda$. 
For the QD, $x_{\rm QD}\sim10^{-4}$ and, therefore, we can realistically assume $\epsilon_{\rm QD}=0$.
At the same time, $x_{\rm MP}$ may approach the value of $0.5$ for the microparticles normally used in experiments.
Therefore, $\epsilon_{\rm MP}$ will need to be calculated according to Mie scattering theory.
Instead, to increase the thermal contrast between the microparticle and the QD, we will assume $\epsilon_{\rm MP}=1$.
\\ \indent
In terms of the collection of the charged species by the quantum dot, the largest reduction of the effective surface area will correspond to the case when the fluxes of charged species are orthogonal to the surface of the microparticle which implies $\alpha_{\rm e,i}=1/4$ and
\begin{equation}
J_{\rm e,i}^{\rm QD} = J_{\rm e,i}^{\rm MP}\left(\frac{a_{\rm QD}}{2a_{\rm MP}}\right)^2.
\label{Eq:JeiQD1div4}
\end{equation}
Regarding the neutral component, a reasonable assumption would be that the neutral atoms will arrive the QD surface from the entire plasma-facing hemisphere which would result in $\alpha_{\rm g}=1/2$ and
\begin{equation}
J_{\rm g}^{\rm QD}=\pi a_{\rm QD}^2v_{\rm nth}\frac{p}{k_{\rm B}T_{\rm g}}.
\label{Eq:JgQD1div2}
\end{equation}
\\ \indent
It is difficult to make any assumption on the accommodation coefficients.
Therefore, we assume $\xi_{\rm MP,QD}=1$.
\subsubsection{Equations}
Using the assumptions formulated in sections~\ref{Sec:HF} and \ref{Sec:Surf}, we can write the equations which will be solved to model the thermal behavior of our system in the pulsed plasma conditions:
\begin{strip}
\begin{eqnarray}
C_{\rm MP}m_{\rm MP}\frac{\partial T_{\rm MP}}{\partial t}=
\left\{
\begin{array}{c l}
\begin{aligned}
&J_{\rm i}^{\rm MP}\left(E_{\rm ion}+e\phi\right)+J_{\rm e}^{\rm MP}k_{\rm B}T_{\rm e}-\\
&-J_{\rm g}^{\rm MP}k_{\rm B}\left(T_{\rm MP}-T_{\rm g}\right)-4\pi a_{\rm MP}^2\sigma\left(T_{\rm MP}^4-T_{\rm g}^4\right)
\end{aligned}
& \textrm{in the reignition}\\
&\\
-J_{\rm g}^{\rm MP}k_{\rm B}\left(T_{\rm MP}-T_{\rm g}\right)-4\pi a_{\rm MP}^2\sigma\left(T_{\rm MP}^4-T_{\rm g}^4\right)& \textrm{in the afterglow}
\end{array},
\right.
\label{Eq:TBMPPulse}\\
C_{\rm QD}m_{\rm QD}\frac{\partial T_{\rm QD}}{\partial t}=
\left\{
\begin{array}{c l}
\begin{aligned}
&J_{\rm i}^{\rm QD}\left(E_{\rm ion}+e\phi\right)+J_{\rm e}^{\rm QD}k_{\rm B}T_{\rm e}-\\
&-J_{\rm g}^{\rm QD}k_{\rm B}\left(T_{\rm QD}-T_{\rm g}\right)-J^{\rm QDMP}k_{\rm B}\left(T_{\rm QD}-T_{\rm MP}\right)
\end{aligned}
& \textrm{in the reignition}\\
&\\
-J_{\rm g}^{\rm QD}k_{\rm B}\left(T_{\rm QD}-T_{\rm g}\right)-J^{\rm QDMP}k_{\rm B}\left(T_{\rm QD}-T_{\rm MP}\right)& \textrm{in the afterglow}
\end{array},
\right.
\label{Eq:TBQDPulse}
\end{eqnarray}
\end{strip}
\noindent
where fluxes $J_{\rm e,i,g}^{\rm MP}$ are calculated using the Equations~\eref{Eq:JeMP}, \eref{Eq:JiMP} and \eref{Eq:JgMP}, respectively, fluxes $J_{\rm e,i,g}^{\rm QD}$ are calculated using Equations~\eref{Eq:JeiQD1div4} and \eref{Eq:JgQD1div2}, respectively, and $\phi$ is obtained by solving~\Eref{Eq:Ch} with $\partial \phi/\partial t = 0$.
As initial conditions for the pulsed plasma case, we use the solutions of Equations~\eref{Eq:TBMP} and \eref{Eq:TBQD} with $\partial T_{\rm MP,QD}/\partial t = 0$, with fluxes and potential $\phi$ calculated as stated above, with emissivities $\epsilon_{\rm MP}=1$ and $\epsilon_{\rm QD}=0$ and with accommodation coefficients $\xi_{\rm MP, QD}=1$.
\section{Results}
\label{Sec:Res}
\begin{table}
\begin{tabular}{||l|c||}
\hline
{\bf Parameter}					&{\bf Value}				\\ \hline
$n_{\rm e}$ [m$^{-3}$]			&$1.0\times10^{15}$			\\ \hline
$k_{\rm B}T_{\rm e}$ [eV]			&$1.3$ 					\\ \hline
$T_{\rm g}$ [K]					&$300$					\\ \hline
$M$ [a.m.u.]					&$40$					\\ \hline
$a_{\rm MP}$ [$\mu$m]			&$4.6$ 					\\ \hline
$C_{\rm MP}$ [J kg$^{-1}$K$^{-1}$]	&$1200$~\cite{designerdata}	\\ \hline
$\rho_{\rm MP}$ [kg m$^{-3}$]		&$1500$~\cite{microparticles}	\\ \hline
$a_{\rm QD}$ [nm]				&$3.3$					\\ \hline
$C_{\rm QD}$ [J kg$^{-1}$K${-1}$]	&$490$~\cite{isspcdse}		\\ \hline
$\rho_{\rm QD}$ [kg m$^{-3}$]		&$5810$\cite{isspcdse}		\\ \hline
\end{tabular}	
\caption{
Input parameters for the calculations of the thermal balance of the QD on the microparticle surface.
The plasma is supposed to be produced in argon gas.
The material of the microparticle is supposed to be melamineformaldehyde, whereas the material of the QD is supposed to be cadmium selenide.
Masses of the microparticle and the QD are calculated as $m_{\rm MP, QD}=\frac{4}{3}\pi\rho_{\rm MP, QD}a_{\rm MP, QD}^3$, where $\rho_{\rm MP, QD}$ are the respective material densities.
For simplicity, we neglect the mass of the QD shell which typically has a thickness of about $0.7$~nm.}
\label{Tab:SummPar}
\end{table}
\begin{figure}[t!]
\includegraphics[width=8cm]{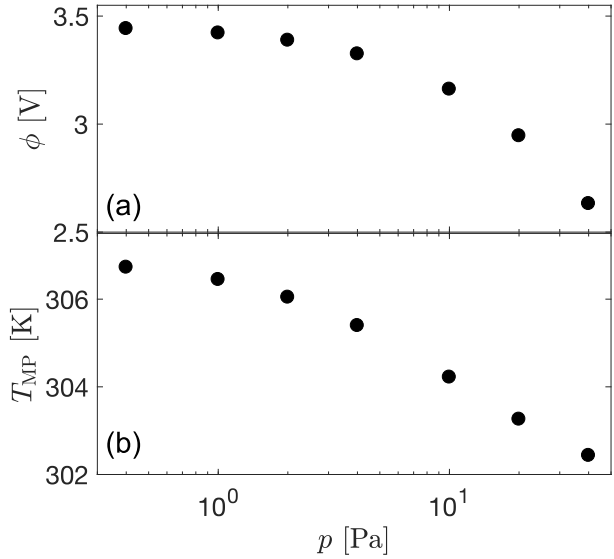}
\caption{Neutral gas pressure dependence of (a) microparticle surface potential $\phi$ and (b) microparticle temperature $T_{\rm MP}$ in steady-state plasma conditions.
The former drops with pressure due to the increasing role of the ion-neutral collisions, whereas the latter - due to the increase of the neutral gas flux to the microparticle surface.
Potential $\phi$ is a steady-state solution of \Eref{Eq:Ch}.
Microparticle temperature is a steady-state solution of \Eref{Eq:TBMP} with the electron and ion fluxes given by \Eref{Eq:JeiQD1div4}, neutral gas flux given by \Eref{Eq:JgQD1div2}, $\epsilon_{\rm MP}=1$, $\epsilon_{\rm QD}=0$ and $\xi_{\rm MP, QD}=1$}
\label{Fig:pdep}
\end{figure}
Calculations are performed for the parameters listed in \Tref{Tab:SummPar}.
The plasma parameters as well as the microparticle size are typical for complex plasma experiments \cite{Nosenko2018}.
Also, in the calculations, we assumed the period of the plasma pulsing of $2\tau=2$~ms with equal afterglow and reignition durations.
We varied argon pressure in the range between $0.4$ to $40$~Pa and the flux $J^{\rm QDMP}$ between $0$ and $10^9$~s$^{-1}$.
For each value of $p$ and $J^{\rm QDMP}$, first, the values of $T_{\rm MP, QD}$ are calculated for steady-state plasmas. 
These values are used as initial conditions for the solution of the pulsed-plasma problem.
\\ \indent
\Fref{Fig:pdep}(a) shows the pressure dependence of the microparticle surface potential in steady-state plasmas.
Its decrease with pressure is due to the well-known effect of trapped ions \cite{Zobnin2000, Lampe2003}.
We note, that the surface potential calculated here is about a factor of three smaller than that measured in \cite{Nosenko2018} since in the experiment, the microparticles levitate in the sheath where the ions cannot be considered cold.
In our model, we, however, assume the ions to be cold.
In \Fref{Fig:pdep}(b), the pressure dependence of the microparticle temperature in steady-state plasmas is shown. 
The decrease of $T_{\rm MP}$ with pressure is an evident consequence of the growth of neutral-gas flux to the surface of the microparticle. 
\\ \indent
\begin{figure*}[t!]
\includegraphics[width=17cm]{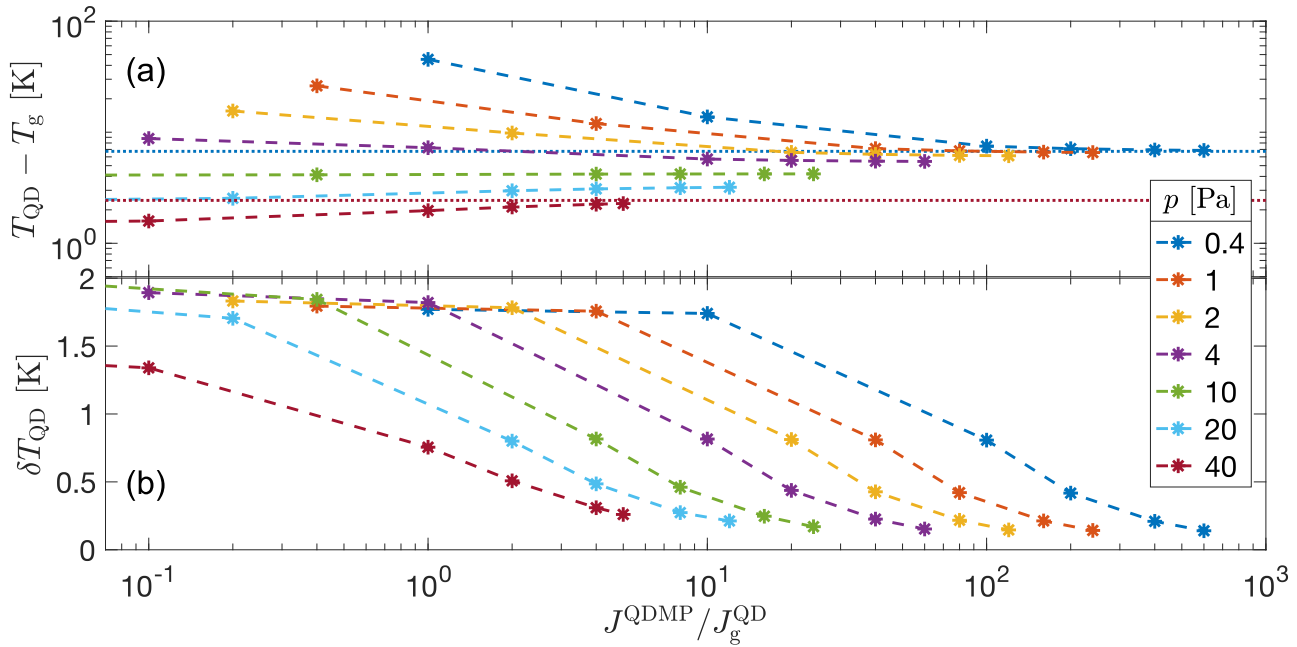}
\caption{Dependence of (a) the QD temperature in steady-state plasma and (b) amplitude of stationary oscillations of the QD temperature in periodically pulsed plasma on the ratio $J^{\rm QDMP}/J_{\rm g}^{\rm QD}$ (dashed lines).
Dotted lines in plate (a) show the values of $T_{\rm MP}-T_{\rm g}$ at $0.4$ and $40$~mbar, respectively.
$T_{\rm g}=300$~K.
QD temperature is a steady-state joint solution of Equations~\eref{Eq:TBMP} and \eref{Eq:TBQD} with the electron and ion fluxes given by \Eref{Eq:JeiQD1div4}, neutral gas flux given by \Eref{Eq:JgQD1div2}, $\epsilon_{\rm MP}=1$, $\epsilon_{\rm QD}=0$, $\xi_{\rm MP, QD}=1$ and $\phi$ being the steady-state solution of \Eref{Eq:Ch}.
Amplitude $\delta T_{\rm QD}$ is determined from the solution for periodically pulsed plasma (\Fref{Fig:TQD}).}
\label{Fig:jqdmpdep}
\end{figure*}
Unlike $T_{\rm MP}$, the temperature of the QD in steady-state plasmas, $T_{\rm QD}$, depends not only on the pressure, but also on the thermal contact between the microparticle and the QD.
Therefore, in \Fref{Fig:jqdmpdep}(a), we show its dependence on both variables.
At large values of $J^{\rm QDMP}$, $T_{\rm QD}$ approaches the value of $T_{\rm MP}$ for the respective pressure.
$T_{\rm MP}$ values for $0.4$ and $40$~Pa only are shown for clarity.
At low values of $J^{\rm QDMP}$, $T_{\rm QD}$ significantly differs from the respective $T_{\rm MP}$.
At low pressures the difference reaches tens of K.
We also note that at low pressure, $T_{\rm QD}>T_{\rm MP}$, whereas at high pressures, the situation is opposite.
The boundary between these two regimes is at the pressure of $\sim 10$~Pa.
\\ \indent
Next, we consider solutions for periodically pulsed plasma.
\Fref{Fig:TMP} shows the evolution of the microparticle temperature $T_{\rm MP}$ at $0.4$~Pa (\Fref{Fig:TMP}(a)) and $40$~Pa (\Fref{Fig:TMP}(b)).
\begin{figure}[t!]
\includegraphics[width=8.6cm]{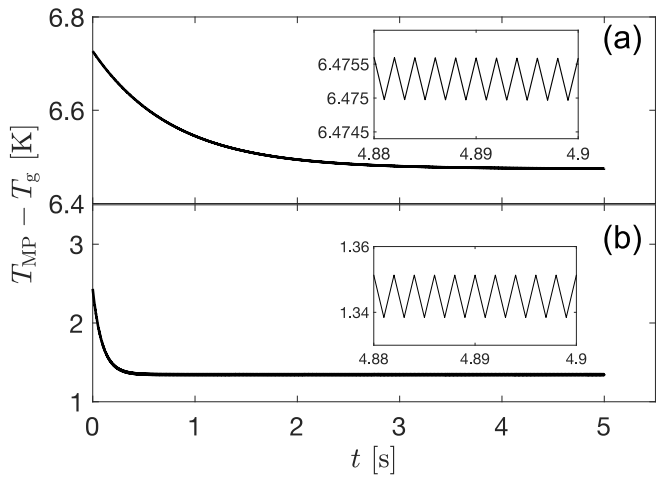}
\caption{Temporal evolution of the microparticle temperature at (a) $p=0.4$~Pa and (b) $p=40$~Pa in periodically pulsed plasma.
At $t=0$, $T_{\rm MP}$ acquires the value for steady-state plasma (\Fref{Fig:pdep}(b)).
$T_{\rm g}=300$~K.
The insets show small stationary temperature oscillations caused by plasma pulsing.
Microparticle temperature for the periodically pulsed plasma is obtained by solving the \Eref{Eq:TBMPPulse} with the electron and ion fluxes given by \Eref{Eq:JeiQD1div4}, neutral gas flux given by \Eref{Eq:JgQD1div2} and $\phi$ given by the steady-state solution of \Eref{Eq:Ch}.}
\label{Fig:TMP}
\end{figure}
With the onset of pulsing, the microparticle temperature decreases due to the decrease of the heat load.
The insets show small stationary oscillations caused by pulsing.
Oscillations are stabilized considerably faster at higher pressure.
Also, the temperature drop at $40$~Pa is much larger ($\approx 1$~K) than that at $0.4$~Pa ($\approx 0.2$~K).
\\ \indent
Evolution of QD temperature $T_{\rm QD}$ in periodically pulsed plasma is shown in \Fref{Fig:TQD} for different values of $J^{\rm QDMP}$ at neutral gas pressures of $0.4$~Pa (\Fref{Fig:TQD}(a)) and $40$~Pa (\Fref{Fig:TQD}(b)).
The insets show the stationary oscillations due to plasma pulsing whose amplitude is much larger than that of $T_{\rm MP}$ due to much smaller thermal capacity of the QD.
We determined the peak-to-peak amplitude of the stationary oscillations $\delta T_{\rm QD}$.
The result is plotted in \Fref{Fig:jqdmpdep}(b) and in \Fref{Fig:jqdmpabsdep}.
Amplitude $\delta T_{\rm QD}$ appears to be rather a weak function of pressure, but, on the contrary, much stronger function of $J^{\rm QDMP}$.
Behavior of $\delta T_{\rm QD}$ with $J^{\rm QDMP}$ is similar at every pressure:
At low values of  $J^{\rm QDMP}$, the decrease of $\delta T_{\rm QD}$ with $J^{\rm QDMP}$ is relatively slow.
It gets significantly faster after $J^{\rm QDMP}\sim 10^7$~s$^{-1}$.
The lower the pressure, the larger is the ratio $J^{\rm QDMP}/J_{\rm g}^{\rm QD}$ at which the trend changes.
\section{Discussion}
\label{Sec:Disc}
\begin{figure*}[t!]
\includegraphics[width=17cm]{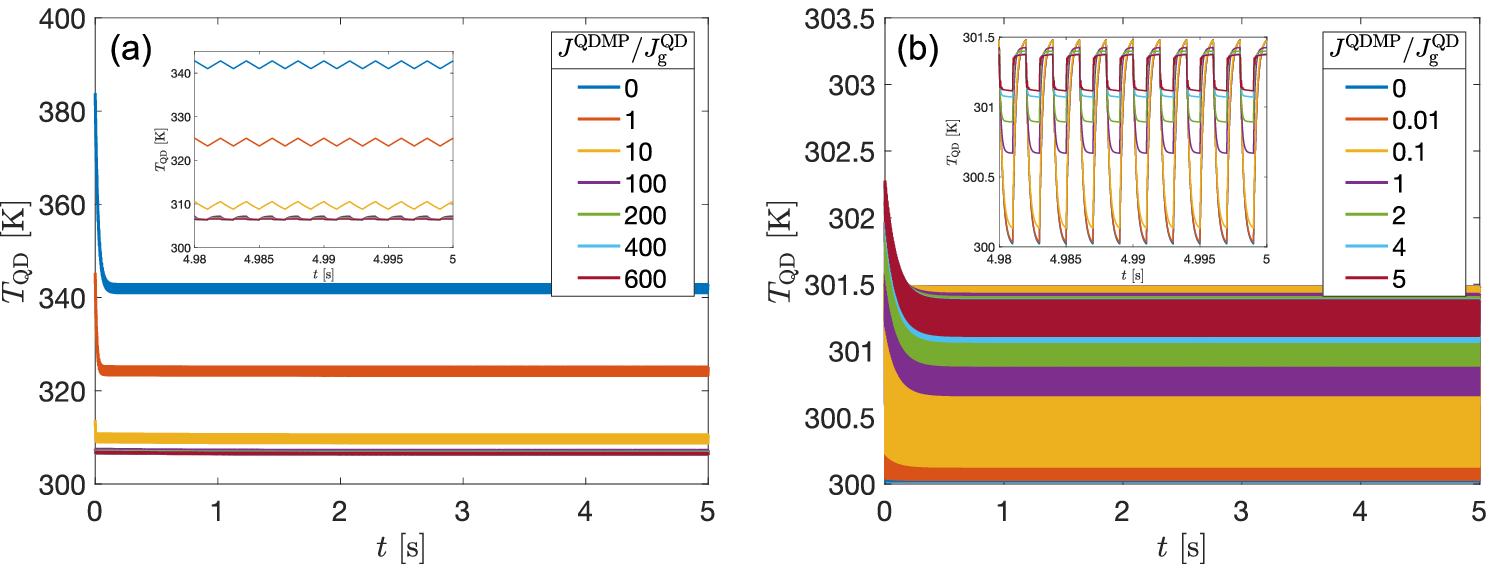}
\caption{Temporal evolution of the QD temperature $T_{\rm QD}$ for different effective fluxes $J^{\rm QDMP}$ at (a) $p=0.4$~Pa and (b) $p=40$~Pa.
$J^{\rm QDMP}$ characterizes the thermal contact between the QD and the microparticle.
At $t=0$, $T_{\rm QD}$ acquires the value for steady-state plasma (\Fref{Fig:jqdmpdep}(a)).
QD temperature for the periodically pulsed plasma is obtained by jointly solving Equations \eref{Eq:TBMPPulse} and \eref{Eq:TBQDPulse} with the electron and ion fluxes given by \Eref{Eq:JeiQD1div4}, neutral gas flux given by \Eref{Eq:JgQD1div2} and $\phi$ given by the steady-state solution of \Eref{Eq:Ch}.
}
\label{Fig:TQD}
\end{figure*}
\subsection{Oscillations of the quantum dot temperature}
\Fref{Fig:TQD} demonstrates the qualitative difference in the character of stationary oscillations of $T_{\rm QD}$ at low and high values of $J^{\rm QDMP}$.
At low $J^{\rm QDMP}$, the oscillations occur between the transient values of $T_{\rm QD}$.
In \Fref{Fig:TQD}(a), they are of almost ideal sawtooth shape with the linear variation in both the afterglow and reignition phases.
In this regime, oscillation amplitude only weakly depends on $J^{\rm QDMP}$ (see \Fref{Fig:jqdmpdep}(b) and \Fref{Fig:jqdmpabsdep}).
However, at large $J^{\rm QDMP}$, the QD temperature oscillates between two \mbox{(quasi-)steady}-state values, one of which is, obviously, only very weakly oscillating $T_{\rm MP}$, and the other is the $T_{\rm QD}$ value for steady-state plasma.
This change in the oscillation waveform is accompanied by strengthening of the dependence of $\delta T_{\rm QD}$ on $J^{\rm QDMP}$.
\\ \indent
Transition between the two oscillation regimes should occur at such $J^{\rm QDMP}$ which would allow the QD temperature variations within the half-period of the oscillations, i.e. $J^{\rm QDMP}\sim \tau^{-1}C_{\rm QD}m_{\rm QD}/k_{\rm B}\approx 3.1\times10^7$~s$^{-1}$.
This result is in accord with the trend change in \Fref{Fig:jqdmpabsdep} at $J^{\rm QDMP}\gtrsim 10^7$~s$^{-1}$.
\subsection{Experimental detectability}
\label{Sec:Thresh}
\begin{figure}[t!]
\includegraphics[width=8.6cm]{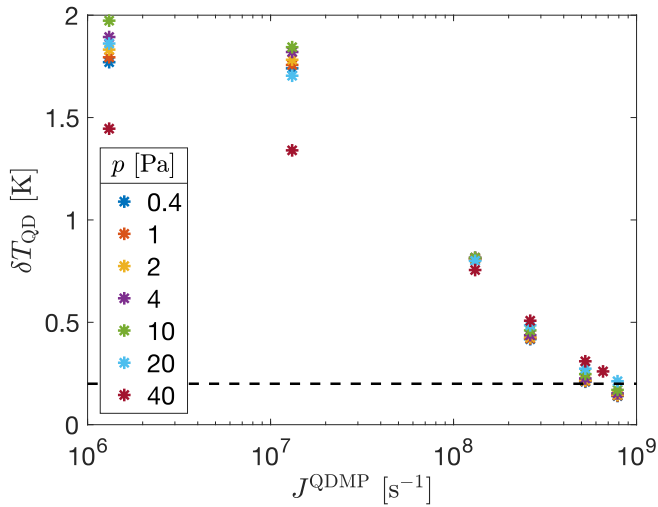}
\caption{Dependence of the amplitude of stationary oscillations of the QD in periodically pulsed plasma on the absolute value of the effective flux $J^{\rm QDMP}$ characterizing the thermal contact between the QD and the microparticle.
The dashed line shows the detectability threshold of $0.2$~K (see \Sref{Sec:Thresh}).}
\label{Fig:jqdmpabsdep}
\end{figure}

In \cite{Pustylnik2021}, the detection limit of the red shift of the QD photoluminescence was determined based on the data of \cite{Marvi2021} as $0.02$~nm.
According to \cite{Walker2003, Valerini2005, Albahrani2018}, the thermal red shift of the CdSe QD photoluminescence is $\sim 0.1$~nm~K$^{-1}$.
Therefore, in the following, we will consider $\delta T_{\rm QD}\lesssim0.2$~K to be undetectable.
In this sense, the oscillations of the microparticle temperature (\Fref{Fig:TMP}) are neglighible, whereas the oscillations of the QD temperature tend to the detection threshold value of $0.2$~K at fluxes $J^{\rm QDMP}\gtrsim 10^9$~s$^{-1}$ (\Fref{Fig:jqdmpabsdep}).
\\ \indent
\subsection{Cooling regimes}
Large oscillations of the temperature of the QD are the consequence of its small thermal capacity.
Thermal contact to the microparticle provides the QD with a thermal connection to a body with much larger thermal capacity so that temperature variation get limited by the thermal resistance of this connection.
From the point of view of the microsensor design, it is, however, important to understand which physical processes determine the requirement for $J^{\rm QDMP}$.
\\ \indent
\Fref{Fig:jqdmpdep} clearly shows that the situation is completely different at low and high pressures:
The oscillation amplitude $\delta T_{\rm QD}$ vanishes at very different values of the ratio $J^{\rm QDMP}/J_{\rm g}^{\rm QD}$ (\Fref{Fig:jqdmpdep}(b)).
E.g., at $40$~Pa, this happens at $J^{\rm QDMP}/J_{\rm g}^{\rm QD}\sim 1$, which means that in this case, the thermal contact between the QD and the microparticle has to compensate mainly the neutral-gas cooling of the QD for the temperature oscillations to vanish.
At $0.4$~Pa, this happens at much higher $J^{\rm QDMP}/J_{\rm g}^{\rm QD}\sim 10^2$.
Therefore, at low pressure, the thermal contact has to compensate the cooling by radiation pressure which becomes the main cooling mechanism.
\\ \indent
This is in accord with the behavior of the temperature $T_{\rm QD}$ in steady-state plasma (\Fref{Fig:jqdmpdep}(a)). 
In absence of the thermal contact term, all the terms in the right-hand sides of Equations \eref{Eq:TBMP} and \eref{Eq:TBQD} are proportional to $a_{\rm MP, QD}$, respectively, which means that there is no explicit dependence of the equilibrium temperature of a plasma-facing spherical object on its size.
Difference in equilibrium values of $T_{\rm MP}$ and $T_{\rm QD}$ originates from (i) the shadowing of the fluxes onto the QD by the microparticle and consequent differences in the coefficients $\alpha_{\rm e,i,g}$, (ii) difference in the accommodation coefficients $\xi_{\rm MP,QD}$ and (iii) difference in the emissivities $\epsilon_{\rm MP,QD}$.
In \Sref{Sec:PPP}, we assumed certain (realistic) relations for the above mentioned coefficients.
At low pressures, where the thermal-radiation cooling dominates over the neutral-gas cooling, $T_{\rm QD}>T_{\rm MP}$ since we neglected thermal radiation for the QD.
At high pressures, on the contrary, the neutral-gas cooling dominates over the thermal-radiation cooling.
We assumed $\xi_{\rm MP,QD}=1$.
Therefore, the accommodation coefficients do not contribute to the difference between the steady-state-plasma values of $T_{\rm QD}$ and $T_{\rm MP}$.
We also assumed $\alpha_{\rm e,i}<\alpha_{\rm g}$.
We, therefore, reduced the effective areas of electron and ion collection more than that of the neutral gas atoms.
This unavoidably lead to the QD being colder than the microparticle in the steady-state plasma.
\\ \indent
We would like to emphasize that although we made realistic assumptions for the coefficients $\alpha_{\rm e,i,g}$, the relation between the accommodation coefficients $\xi_{\rm MP,QD}$ is unknown.
Therefore, our conclusion that at high pressures, in steady-state plasma, $T_{\rm QD}<T_{\rm MP}$ cannot pretend for universality.
\subsection{Thermal contact}
Exact calculation of $J^{\rm QDMP}$ in real conditions is connected with many uncertainties and represents, in general, a difficult task.
For estimation, let us assume that the thermal contact between the QD and the microparticle is provided by a pair of atoms one of which belongs to the QD and the other belongs to the microparticle. 
In this case, $J^{\rm QDMP}\sim\sqrt{\frac{\mu_1}{\mu_2}}\nu_{\rm E}$, where $\mu_1/\mu_2<1$ is the ratio of the average masses of the atoms of the microparticle and QD materials, and $\nu_{\rm E}$ is the Einstein frequency of the hotter material.
According to \cite{Freik2014}, the Einstein temperature $T_{\rm E}$ for CdSe is of the order of $100$~K.
Frequency $\nu_{\rm E}=k_{\rm B}T_{\rm E}/h$ and $J^{\rm QDMP}\sim 10^{12}$~s$^{-1}$.
This value is three orders of magnitude larger than what is according to \Sref{Sec:Thresh} required to reduce the temperature oscillations in periodically pulsed plasma below the detectability threshold.
\\ \indent
This means that in the situation of \cite{Marvi2021}, where the QDs were deposited on the substrate in vacuum {\it in situ} without any chemical ligands, the thermal issue should not play any role.
However, the necessity to use (especially, long) ligand molecules to attach the QDs to the microparticle surface might require a more detailed consideration of the thermal connection between the microparticle and the QD. 
\\ \indent
In \cite{Pustylnik2021}, the usage of protective coatings on top of the QDs was suggested.
Also, we did not take into account the QD shell when calculating its thermal capacity (see \Tref{Tab:SummPar}). 
Both of these complications can only increase the effective heat capacity of each QD in the sensitive layer and therefore decrease the QD temperature oscillations.
In addition to that, we note that throughout the paper, we used, where possible, reasonable assumptions that lead to the increase of the QD temperature oscillations.
\section{Conclusion}
\label{Sec:Conc}
We have considered thermal balance of the microparticle and of the quantum dot residing on the plasma-facing surface of the microparticle under steady-state and periodically pulsed plasma conditions.
This problem is important in the scope of measurement of microparticle charges using quantum dots \cite{Pustylnik2021} since both heating and charging cause comparable spectral shifts of the quantum dot photoluminescence.
Thermal and Stark spectral shifts can be distinguished by pulsing the plasma due to the large gap in the timescales between microparticle charging and heating, but only under the condition of sufficiently strong thermal contact between the quantum dot and the microparticle.
\\ \indent
In our thermal model, we assumed that the plasma-facing surfaces are heated by the fluxes of charged particles and cooled down by the neutral gas flux and thermal radiation.
We assumed orbit-motion-limited flux for electrons, whereas, for ions, we took into account the collisions leading to ion trapping.
For pulsed plasmas, we assumed absence of heating fluxes in the afterglow phase.
We calculated the amplitude of the quantum dot temperature oscillations in pulsed isotropic plasmas depending on the neutral gas pressure and on the effective flux characterizing the thermal contact between the quantum dot and the microparticle at typical plasma density and electron density.
\\ \indent
We found this amplitude to be a very weak function of gas pressure.
It exhibited a much stronger dependence on the thermal contact flux.
While the thermal contact was weak, the oscillations occurred between the transient values of temperature.
As the effective flux exceeded the value of $\gtrsim 10^7$~s$^{-1}$, the character of the oscillations changed: The temperature then started to oscillate between the quantum dot temperature in steady-state plasma and the temperature of the microparticle.
These two temperatures differ due to the combination of differences in the surface conditions on the two bodies.
At low pressures, the quantum dots are significantly hotter than the microparticles due to suppression of thermal radiation from the quantum dots.
At high pressures, differences in the accommodation coefficients for neutral gas atoms on the two surfaces, as well differences in effective collection areas for the charged species and neutral gas atoms on the quantum dot play the major role.
\\ \indent
At the values of the effective flux $\sim 10^9$~s$^{-1}$, the quantum dot temperature oscillation amplitude reduces to the values of about $0.2$~K that would be experimentally undetectable.
Under these conditions, the entire spectral shift observed during the period of plasma pulsing should be attributed to the quantum-confined Stark effect due to the microparticle charge.
Under the assumption that the thermal contact is provided by only a pair of atoms, - one belonging to the quantum dot and the other to the microparticle, - the effective flux is estimated to be of the order of Einstein frequency of one of the materials, i.e. $\sim 10^{12}$~s$^{-1}$ which is much larger than the requirement imposed by the heat exchange with the plasma.
Shells of the quantum dots as well as usage of protective layers increase the effective thermal capacity of each quantum dot and, therefore, reduce the temperature oscillations even further.
However, in case of attachment of the quantum dots to the surface with the help of long ligands, the thermal issue should be considered in more details.
\section{Acknowledgements}
The author thanks G. Klaassen, S. Hasani, Prof. J. Beckers and Dr. H. Thomas for careful reading of the manuscript and for helpful discussions.\\
\bibliography{QDThermalBalance.bib}

\end{document}